\documentclass[aps,prd,preprint,amsmath,amssymb,latexsym,superscriptaddress]{revtex4}
\usepackage{graphicx}
\usepackage{dcolumn}
\usepackage{color}
\allowdisplaybreaks

\begin{document}
\title{Increasing the sensitivity of future gravitational-wave detectors with double squeezed-input}
\author{Farid Ya. Khalili}
\affiliation{Physics Faculty, Moscow State University, Moscow 119992, Russia}
\author{Haixing Miao}
\affiliation{School of Physics, University of Western Australia, WA 6009, Australia}
\date{\today}
\author{Yanbei Chen}
\affiliation{Theoretical Astrophysics 130-33, California Institute of Technology,
Pasadena, CA 91125, USA}
\affiliation{Max-Planck Institut f\"ur Gravitationsphysik (Albert-Einstein-Institut),
Am M\"uhlenberg 1, 14476 Golm, Germany}

\newcommand{\be}{\begin{equation}}
\newcommand{\ee}{\end{equation}}
\newcommand{\ba}{\begin{eqnarray}}
\newcommand{\ea}{\end{eqnarray}}

\begin{abstract}

We consider improving the sensitivity of future interferometric gravitational-wave detectors by
simultaneously injecting two squeezed vacuums (light), filtered through a resonant Fabry-Perot cavity, into
the dark port of the interferometer.
 The same scheme with single squeezed vacuum was first proposed
and analyzed by Corbitt {\it et al.} \cite{Corbitt}. Here we show that the extra squeezed vacuum,
together with an additional homodyne detection suggested previously by one of the authors \cite{Khalili1},
allows reduction of quantum noise over the entire detection band. To motivate future implementations,
we take into account a realistic technical noise budget for Advanced LIGO (AdvLIGO) and numerically
optimize the parameters of both the filter and the interferometer for detecting gravitational-wave
signals from two important astrophysics sources, namely Neutron-Star--Neutron-Star (NSNS) binaries
and Bursts. Assuming the optical loss of the $\sim 30$m filter cavity to be $10$ppm per bounce and 10dB
squeezing injection, the corresponding quantum noise with optimal parameters lowers by a factor of 10
at high frequencies and goes below the technical noise at low and intermediate frequencies.
\end{abstract}
\pacs{}

\maketitle

\section{Introduction}

During the last decade, several laser interferometric gravitational-wave detectors including LIGO
\cite{LIGO}, VIRGO \cite{VIRGO}, GEO600 \cite{GEO} and TAMA \cite{TAMA} have been built and operated
almost at their design sensitivity, aiming at extracting gravitational-wave signals from
various astrophysical sources. At present, developments of next-generation detectors such as
AdvLIGO \cite{AdvLIGO} are also under way, and sensitivities of these advanced detectors are
anticipated to be limited by quantum noise nearly over the whole observational band from $10$ Hz
to $10^4$ Hz. At high frequencies, the dominant quantum noise is \textit{photon shot noise},
caused by phase fluctuation of the optical field; while at low frequencies, the
\textit{radiation-pressure noise}, due to the amplitude fluctuation, dominates and it exerts a
noisy random force on the probe masses. These two noises, if uncorrelated, will impose a lower
bound on the noise spectrum, which is called the Standard Quantum Limit (SQL).
In terms of gravitational-wave strain $h\equiv \Delta L/L$, it is given by
\be
S_h^{\rm SQL}=\frac{8\hbar}{m\,\Omega^2 L^2}.
\ee
It can also be derived from the fact that position measurements of the free test mass
do not commute with themselves at different time \cite{Thorne}.

\begin{figure}[!h]
\centerline{\includegraphics[width=0.45\textwidth,bb=20 14 270 280,clip]{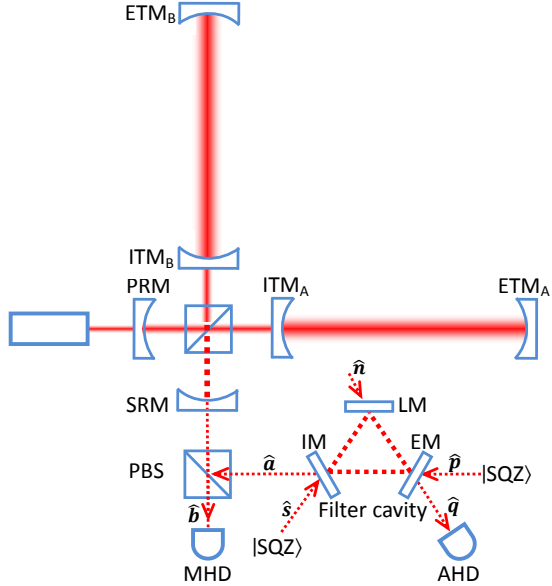}}\caption{
\label{configuration} Schematic plot of the proposed configuration. Two squeezed vacuums $
\hat{\textbf{\textit{s}}}, \hat{\textbf{\textit{p}}}$ are
injected from both side of the filter cavity rather than the one which was considered in
Refs. \cite{Corbitt,Khalili1}. The signal is detected by the main homodyne detector
({\sf MHD}) and additional detector ({\sf AHD}) is made in the idle port of the filter cavity.}
\end{figure}

The existence of SQL was first realized by Braginsky in 1960's \cite{Braginsky1,Braginsky2}.
Since then, various approaches are proposed to beat the SQL. One recognized by Braginsky is to
measure conserved quantities of the probe masses (also called quantum nondemolition quantities).
This can be achieved, e.g. by adopting speed-meter configurations \cite{Khalili2,Braginsky3,
Purdue1,Purdue2,Yanbei,Danilishin}, which measure the conserved quantity -- momentum rather than
the position. An alternative is to change the dynamics of the probe mass, e.g. using optical
rigidity \cite{Braginsky4,Khalili3}, in which case the free mass SQL mentioned is no longer
relevant. As shown by Buonanno and Chen  \cite{Buonanno1, Buonanno2, Buonanno3}, optical rigidity
exists in signal-recycled (SR) interferometric gravitational-wave detectors; therefore we can beat
the SQL without radical redesigns of existing topology of the interferometers. Another approach
is to modify input and/or output optics of the interferometers such that {\it photon shot noise}
and {\it radiation-pressure noise} are correlated. After the initial paper by Unruh \cite{Unruh1982},
this was further developed by other authors \cite{87a1eKh, JaekelReynaud1990, Pace1993, Vyatchanin,
Kimble, Buonanno4, Corbitt, Arcizet2006, 06a2Kh}. A natural way to achieve this is injecting
squeezed vacuum, whose phase and amplitude fluctuations are correlated, into the dark port of the
interferometers. With great advancements in preparation of the squeezed state \cite{McKenzie,Vahlbruch},
squeezed-input interferometers will be promising candidates for third-generation gravitational-wave
detectors. As elaborated in the work of Kimble \textit{et al.} \cite{Kimble}, frequency-dependent
squeezing is essential to reduce the quantum noise at various frequencies of the observation band.
In addition, they demonstrated that this can be realized by filtering the frequency-independent
squeezed vacuum through two detuned Fabry-Perot cavities before sending into the interferometer.
Their results were extended by Buonanno and Chen \cite{Buonanno4} where the filters for general
cases were discussed.

Another method, which also uses an additional filter cavity and squeezed light, but in a completely different way,
was proposed by Corbitt, Mavalvala, and Whitcomb (hereafter  referred as CMW) \cite{Corbitt}. They proposed to
use a tuned optical cavity as a high-pass filter for the squeezed vacuum. This scheme does not create the noises
correlation, but instead, renders the noises spectral densities frequency dependent. At high frequencies, the
phase squeezed vacuum gets reflected by the filter
and enters the interferometer such that high-frequency shot noise is reduced; while at low frequencies,
ordinary vacuum transmits through the filter and enters the interferometer, thus low-frequency
radiation-pressure noise remains unchanged. One significant advantage is that the squeezed
vacuum does not really enter the filter cavity and thus it is less susceptible to the optical losses.
However, it does not perform so well as hoped for, and there is a noticeable degradation
of sensitivity in the intermediate-frequency range. One of us -- Khalili \cite{Khalili1} pointed
out that this has to do with the quantum entanglement between the optical fields at two ports of
the filter cavity. Equivalently, it can be interpreted physically as the following: some information
about phase and amplitude fluctuation flows out from the idle port of the filter cavity and the
remaining quantum state which enters the interferometer is not pure. In order to recover the
sensitivity, the filter cavity needs to have a low optical loss such that these information can
be collected with an additional homodyne detector ({\sf AHD}) at the idle port. Given an achievable
optical loss of the filter cavity $\sim 10$ppm per bounce, Khalili showed that we can obtain the desired
sensitivity at intermediate frequencies. A natural extension of this scheme is sending additional
squeezed vacuum into the idle port of the filter cavity such that the low-frequencies radiation-pressure noise is also
suppressed. The corresponding configuration is shown schematically in Fig. \ref{configuration}, where
two squeezed vacuums $\hat{\textbf{\textit{s}}}$ and $\hat{\textbf{\textit{p}}}$ are injected from
two ports of the filter cavity, and some ordinary vacuum $\hat{\textbf{\textit{n}}}$ leaks into
the filter due to optical losses. By optimizing the squeezing angles of squeezed states, we will show
that the resulting quantum noise is reduced over the entire observational band.

The outline of this paper is as follows. In Sec. \ref{II}, we will calculate the quantum noises
in this double squeezed-input CMW scheme with {\sf AHD} (later referred as CMWA). We will use the same
notation as in Ref \cite{Buonanno3}, which enables us to extend the
results in Ref. \cite{Khalili1} to the case of signal-recycled interferometers easily. In Sec.
\ref{III}, we numerically optimize the parameters of this new scheme for searching the gravitational-wave signals
from NSNS binaries and Bursts. Finally, we will summarize our results in Sec. \ref{IV}. For
simplicity, we will neglect the optical losses inside the main interferometer, but we consider
the losses from the filter cavity and also non-unity quantum efficiency
of the photodiodes. The losses from the main interferometer are not expected to be important
as shown in Refs. \cite{Kimble,Buonanno1,Khalili4}. The main notations used in this paper are listed in Table \ref{tab:notations}.

\begin{table}[!h]
\renewcommand{\arraystretch}{0.8}
 \caption{Main notations used in this paper.}\label{tab:notations}
  \begin{tabular}{lp{4 cm}ll}

    \hline
    \hline
      Quantity           &Value for Estimates                 & Descriptions \\
    \hline
      $\Omega$           &                                    &Gravitational wave (sideband) frequency \\
      $c$                & $3.0\times 10^8 $~m/s                & Speed of light \\
      $\omega_0$         & $1.8\times 10^{15}~{\rm s}^{-1}$  & Optical pumping frequency \\
      $m$                & 40~kg                              & Mass of the end mirrors \\
      $L$                & 4~km                               & Length of the arm cavity\\
      $I_c$              & 840~kW                             & Circulating optical power\\
      $\iota_c=\frac{8 \omega_0 I_c}{m L c}$ & $(2\pi\times 100)^3 {\rm s}^{-3}$ &\\[1ex]
      $\gamma_{\rm arm}$ & $2\pi\times 100~{\rm s}^{-1}$      & Bandwidth of the arm cavity\\
      $r_{\rm SR}$             &                                    & Amplitude reflectivity of the {\sf SRM}\\
      $\phi_{\rm SR}$    &                                    & Phase detuning of the SR cavity\\
      $\delta$           &                                    & Effective frequency detuning of the SR interferometer\\
      $\gamma$           &                                    & Effective bandwidth of the SR interferometer\\
      $\phi$             &                                    & Homodyne angle of the {\sf MHD}\\
      $\zeta$            &                                    & Homodyne angle of the {\sf AHD}\\
      $L_f$              &30~m                                & Length of the filter cavity\\[1ex]
      $\gamma_{\rm\sf I,E,L} = \frac{T_{\rm\sf I,E,L}^2}{2\tau_f}$& &\\[1ex]
      $\gamma_f = \gamma_{\rm\sf I}+\gamma_{\rm\sf E}+\gamma_{\rm\sf L}$ & & Bandwidth of the filter cavity\\[1ex]
      $r_i~(i=s,p)$      & $(\ln 10)/2\;$ (10dB)                      & Squeezing factors \\
      $\theta_i~(i=s,p)$ &                                    & Squeezing angles \\
    \hline
  \end{tabular}
\end{table}
\newpage

\newpage
\section{Quantum noise calculation}\label{II}
\subsection{Filter cavity}

In this section, we will derive single-sided spectral densities of the two
outgoing fields $\hat{\textbf{\textit{a}}}$ and $\hat{\textbf{\textit{q}}}$ as
shown in Fig. \ref{configuration}. From the continuity of optical fields, we can relate them
to ingoing fields, which include two squeezed vacuums $\hat{\textbf{\textit{s}}},~\hat{\textbf{\textit{p}}}$
and one ordinary vacuum $\hat{\textbf{\textit{n}}}$ entering from the lossy mirror (\textsf{LM}) \cite{Khalili1}.
Specifically, we have
\begin{eqnarray}
\label{aq}
\hat{\textbf{\textit{a}}}(\Omega)&=&{\cal R}_{\rm\sf I}(\Omega)\hat{\textbf{\textit{s}}}(\Omega)+
{\cal T}(\Omega)\hat{\textbf{\textit{p}}}(\Omega)+{\cal A}_{\rm\sf I}(\Omega)\hat{\textbf{\textit{n}}}(\Omega),\\
\hat{\textbf{\textit{q}}}(\Omega)&=&{\cal R}_{\rm\sf E}(\Omega)\hat{\textbf{\textit{p}}}(\Omega)+
{\cal T}(\Omega)\hat{\textbf{\textit{s}}}(\Omega)+{\cal A}_{\rm\sf E}(\Omega)\hat{\textbf{\textit{n}}}(\Omega).
\end{eqnarray}
Here $\hat{\textbf{\textit{a}}}=(\hat{a}_{A},~\hat{a}_{\varphi})^{\rm T}$,
$\hat{\textbf{\textit{q}}}=(\hat{q}_{A},~\hat{q}_{\varphi})^{\rm T}$  are amplitude and phase quadratures (subscript
$A$ stands for amplitude and $\varphi$ for phase). In our case, the carrier light is resonant inside the filter cavity.
Therefore, the effective amplitude reflectivity ${\cal R}$, transmissivity ${\cal T}$ and loss ${\cal A}$ can be written as,
\begin{subequations}
  \begin{align}
    {\cal R}_{\rm\sf I}(\Omega) &= \frac{\gamma_{\rm\sf I} - \gamma_{\rm\sf E} - \gamma_{\rm\sf L} + i\Omega}
      {\gamma_{f} - i\Omega} \,, &
    {\cal R}_{\rm\sf E}(\Omega) &= \frac{\gamma_{\rm\sf E} - \gamma_{\rm\sf I} - \gamma_{\rm\sf L} + i\Omega}
      {\gamma_{f} - i\Omega} \,, \\
      {\cal T}(\Omega) &= \frac{-2\sqrt{\gamma_{\rm\sf I}\gamma_{\rm\sf E}}}{\gamma_{f} - i\Omega},\\
    {\cal A}_{\rm\sf I}(\Omega) &= \frac{-2\sqrt{\gamma_{\rm\sf I}\gamma_{\rm\sf L}}}{\gamma_{f} - i\Omega}  \,, &
    {\cal A}_{\rm\sf E}(\Omega) &= \frac{2\sqrt{\gamma_{\rm\sf E}\gamma_{\rm\sf L}}}{\gamma_{f} - i\Omega}  \,
  \end{align}
\end{subequations}
where $\gamma_{f}\equiv\gamma_{\rm\sf I}+\gamma_{\rm\sf E}+\gamma_{\rm\sf L}$.
They satisfy the following identities,
\begin{subequations}
  \begin{align}
    &|{\cal R}_{\rm\sf I}(\Omega)|^2+|{\cal T}(\Omega)|^2+|{\cal A}_{\rm\sf I}(\Omega)|^2=
    |{\cal R}_{\rm\sf E}(\Omega)|^2+|{\cal T}(\Omega)|^2+|{\cal A}_{\rm\sf E}(\Omega)|^2=1,\\
    &{\cal R}^*_{\rm\sf I}(\Omega){\cal T}(\Omega)+{\cal R}_{\rm\sf E}(\Omega){\cal T}^*(\Omega)+
    {\cal A}^*_{\rm\sf I}(\Omega){\cal A}_{\rm\sf E}(\Omega)=0.
  \end{align}
\end{subequations}
If the input and end mirrors of the filter cavity are identical, namely $\gamma_{\rm\sf I}=\gamma_{\rm\sf E}$,
we will have ${\cal R}_{\rm\sf I}, {\cal R}_{\rm\sf E}\sim 1$ and ${\cal T}\sim 0$ when $\Omega\gg \gamma_f$
and ${\cal R}_{\rm\sf I}, {\cal R}_{\rm\sf E}\sim 0$ and ${\cal T}\sim 1$ when $\Omega\ll \gamma_f$.
Therefore, the squeezed vacuum $\hat{\textbf{\textit{s}}}$ enters the interferometer at high frequencies
while $\hat{\textbf{\textit{p}}}$ becomes significant mostly at low frequencies. By adjusting the squeezing factor and angle of these two squeezed fields,
we can reduce both the high-frequency shot noise and low-frequency radiation-pressure noise simultaneously.

To calculate the noise spectral densities, we assume these two squeezed vacuums have frequency-independent squeezing angles $\theta_i~(i=s,~p)$ and
can be represented as follows:
\begin{align}
  \hat{\textbf{\textit{s}}} &= \widetilde{R}(r_s,\theta_s)\hat{\textbf{\textit{v}}}_s \,, &
  \hat{\textbf{\textit{p}}} &= \widetilde{R}(r_p,\theta_p)\hat{\textbf{\textit{v}}}_p  \,,
\end{align}
with
\be
\widetilde{R}(r,\theta)\equiv\left(\begin{array}{cc}
\cosh r-\cos\theta\sinh r&-\sin\theta\sinh r
\\-\sin\theta\sinh r
&\cosh r+\cos \theta\sinh r\end{array}\right).
\ee
Here $r$ is the squeezing factor, and $\hat{\textbf{\textit{v}}}_i=(\hat{v}_{iA},~\hat{v}_{i\varphi})^{\rm T}$ are ordinary
vacuums with single-sided spectral densities:
$S_{A}(\Omega)=S_{\varphi}(\Omega)=1,~~S_{A\varphi}(\Omega)=0$ \cite{Kimble}.

The noise spectral densities of the two filter-cavity outputs can then be written as
\ba
\widetilde{S}_{\hat a\hat a}(\Omega)&=&|{\cal R}_{\rm\sf I}(\Omega)|^2 \widetilde{R}(2r_s,2\theta_s)
+|{\cal T}(\Omega)|^2\widetilde{R}(2r_p,2\theta_p)+|{\cal A}_{\rm\sf I}(\Omega)|^2\widetilde{\rm I},\\
\widetilde{S}_{\hat q\hat q}(\Omega)&=&|{\cal R}_{\rm\sf E}(\Omega)|^2 \widetilde{R}(2r_p,2\theta_p)
+|{\cal T}(\Omega)|^2\widetilde{R}(2r_s,2\theta_s)+|{\cal A}_{\rm\sf E}(\Omega)|^2\widetilde{\rm I},\\
\widetilde{S}_{\hat a \hat q}(\Omega)&=&{\cal R}_{\rm\sf E}(\Omega){\cal T}^*(\Omega) \widetilde{R}(2r_p,2\theta_p)
+{\cal R}_{\rm\sf I}^*(\Omega){\cal T}(\Omega)\widetilde{R}(2r_s,2\theta_s)+{\cal A}_{\rm\sf E}(\Omega)
{\cal A}_{\rm\sf I}^*(\Omega)\widetilde{\rm I},
\ea
where $\widetilde{S}_{\hat a \hat q}(\Omega)$ is the cross
correlation between two outputs; $\widetilde{\rm I}$ is the identity matrix and
\be
\widetilde{S}_{i}(\Omega)=\left(\begin{array}{cc}S_{A,{i}}(\Omega)&S_{A\varphi,{i}}(\Omega)
\\S_{A\varphi,{i}}(\Omega)
&S_{\varphi,{i}}(\Omega)\end{array}\right)
\ee
with $i=\hat a\hat a,\hat q\hat q,\hat a\hat q$, whose elements are single-sided spectral densities.

\subsection{Quantum noise of the interferometer}
According to Ref. \cite{Buonanno3}, the input-output relation, which connects ingoing fields
$\hat{\textbf{\textit{a}}}$ and gravitational-wave signal $h$ with the outgoing fields $\hat{\textbf{\textit{b}}}$,
for a signal-recycled interferometer can be written as
\begin{equation}
\hat{\textbf{\textit{b}}}=\frac{1}{M}\left(\widetilde{C}~\hat{\textbf{\textit{a}}}+{\bf D}~\frac{h}{h_{\rm SQL}}\right).
\end{equation}
In the above equation,
\begin{equation}
M=[\delta^2-(\Omega+i\gamma)^2]\Omega^2-\delta\,\iota_c,
\end{equation}
and
$\widetilde{C}$ is the transfer function matrix with elements
\begin{subequations}
\begin{align}
\widetilde{C}_{11}&=\widetilde{C}_{22}=\Omega^2(\Omega^2-\delta^2+\gamma^2)+\delta\,\iota_c,\\
\widetilde{C}_{12}&=-2\gamma\,\delta\,\Omega^2,~~~
\widetilde{C}_{21}=2\gamma\,\delta\,\Omega^2-2\gamma\,\iota_c,
\end{align}
\end{subequations}
where $\iota_c\equiv 8\omega_0 I_c/(m\,c\,L)$. The elements of the transfer function vector ${\bf D}$ are
\begin{subequations}
\begin{align}
D_1&=-2\delta\sqrt{\gamma\,\iota_c}\,\Omega,~
D_2=-2(\gamma-i\Omega)\sqrt{\gamma\,\iota_c}\,\Omega.
\end{align}
\end{subequations}
The effective detuning $\delta$ and bandwidth $\gamma$ are given by
\begin{subequations}
\begin{align}
\delta&=\frac{2\,r_{\rm SR}\,\gamma_{\rm arm}\sin(2\,\phi_{\rm SR})}{1+r_{\rm SR}^2+2\,r_{\rm SR}\cos(2\phi_{\rm SR})},\\
\gamma&=\frac{(1-r_{\rm SR}^2)\gamma_{\rm arm}}{1+r_{\rm SR}^2+2\,r_{\rm SR}\cos(2\,\phi_{\rm SR})},
\end{align}
\end{subequations}
where $r_{\rm SR}$ is the amplitude reflectivity of the signal recycling mirror ($\rm\sf SRM$) and $\phi_{\rm SR}$
is phase detuning of the SR cavity.
If the main homodyne detector ({\sf MHD}) measures
\be
\hat{b}_{\phi}(\Omega)=\sqrt{\eta}\,[\sin\phi~\hat{b}_A(\Omega)+\cos\phi~\hat{b}_{\varphi}(\Omega)]+\sqrt{1-\eta}~
\hat{v}(\Omega),
\ee
where $\phi$ is the homodyne angle and $\hat{v}$ is the additional vacuum due to non-unity quantum efficiency of the photodiode,
and then the corresponding $h$-referred quantum-noise spectral density can be written as
\begin{equation}
\label{sh}
S_h(\Omega)=h^2_{\rm SQL}\frac{(\sin\phi~\cos\phi)\widetilde{C}\,\widetilde{S}_{\rm\sf I}\,\widetilde{C}^{\dag}
(\sin\phi~\cos\phi)^{\rm T}+\frac{1-\eta}{\eta}|M|^2}
{(\sin\phi~\cos\phi)\widetilde{D}\,\widetilde{D}^{\dag}(\sin\phi~\cos\phi)^{\rm T}}.
\end{equation}
It can be minimized by adjusting the squeezing angle $\theta$ of $\hat{\textbf{\textit{s}}}$ and
$\hat{\textbf{\textit{p}}}$. We can estimate the optimal $\theta$ qualitatively from the asymptotic behavior
of the resulting noise spectrum. At very high frequencies ($\Omega\gg\gamma$), from Eq. \eqref{aq},
$\hat{\textbf{\textit{a}}}\sim \hat{\textbf{\textit{s}}}$ and thus
\be
S_h(\Omega)\propto\cosh(2r_s)+\cos[2(\phi+\theta_s)]\sinh(2r_s).
\ee
If the squeezing angle of $\hat{\textbf{\textit{s}}}$
\be\label{thetasopt}
\theta_s=\frac{\pi}{2}+n\pi-\phi
\ee
where $n$ is integer, we achieve the optimal case, namely $S_h\propto e^{-2r_s}$.
Similarly, at very low frequencies ($\Omega\ll\gamma$), we have $\hat{\textbf{\textit{a}}}\sim\hat{\textbf{\textit{p}}}$
and the spectral density $S_h\propto e^{-2r_p}$ if
\be\label{thetapopt}
\theta_p=\arctan\left[\frac{2\cos\phi\sin\beta}{\cos(\beta-\phi)+3\cos(\beta+\phi)}\right].
\ee
More accurate values for optimal $\theta_{s,p}$ can be obtained numerically as we will apply in the
next section. Given optimal $\theta_{s,p}$, the sensitivity of this double squeezed-input scheme
improves at both high and low frequencies. However, due to the same reason as in the case of
single squeezed-input that two outputs of the filter cavity are entangled \cite{Khalili1},
this double squeezed-input scheme does not perform well in the intermediate frequency range. To recover the sensitivity,
we need to use an additional homodyne detector ({\sf AHD}) at the idle port E of the filter cavity. The corresponding measured quantity is
\be
\hat{q}_{\zeta}(\Omega)=\sqrt{\eta}\,[\sin\zeta~\hat{b}_A(\Omega)+\cos\zeta~\hat{b}_{\varphi}(\Omega)]+\sqrt{1-\eta}~
\hat{v'}(\Omega)
\ee
where $\zeta$ is homodyne angle and $v'(\Omega)$ is the addition vacuum, which enters due to non-unity
quantum efficiency of photodiode. We combine $\hat{q}_{\zeta}(\Omega)$ with the output
$\hat{b}_{\phi}(\Omega)$ using a linear filter ${\cal K}(\Omega)$, obtaining
\be
\hat{o}(\Omega)=\hat{b}_{\phi}(\Omega)-{\cal K}(\Omega)\,\hat{q}_{\zeta}(\Omega).
\ee
Corresponding, the noise spectral density of this new output $\hat{o}(\Omega)$ can be written as
\be
S_{\hat{o}}(\Omega)=S_{\hat{b}_{\phi}}(\Omega)-2\Re[{\cal K}(\Omega) S_{\hat{b}_{\phi},\hat{q}_{\zeta}}(\Omega)]+|{\cal K}(\Omega)|^2
S_{\hat{q}_{\zeta}}(\Omega).
\ee
The minimum quantum noise is achieved when ${\cal K}(\Omega)=S_{\hat{b}_{\phi},\hat{q}_{\zeta}}(\Omega)/S_{\hat{b}_{\phi}}(\Omega)$ and
the resulting $h$-referred noise spectrum with {\sf AHD} is then
\be
S_h^{\rm\sf AHD}(\Omega)=S_h(\Omega)-h_{\rm SQL}^2\frac{\eta |(\sin\phi~\cos\phi)\widetilde{C}\, \widetilde{S}_{\rm\sf IE}
(\sin\zeta~\cos\zeta)^{\rm T}|^2}{(\sin\phi~\cos\phi)\widetilde{D}\widetilde{D}^{\dag}(
\sin\phi~\cos\phi)^{\rm T}~S_{\zeta}(\Omega)},
\ee
where $S_{\zeta}(\Omega)\equiv \eta(\sin\zeta~\cos\zeta)\widetilde{S}_{\rm\sf E}
(\sin\zeta~\cos\zeta)^{\rm T}+1-\eta$. The second term has a minus sign, which shows explicitly that the sensitivity increases as a result of additional
detection.

\section{Numerical optimizations}\label{III}

In this section, we will take into account realistic technical noise and
numerically optimize interferometer parameters for detecting gravitational-wave signals from specific astrophysics sources which include Neutron-Star-Neutron-Star (NSNS) binaries and Bursts.

For a binaries system, according
to Ref. \cite{Yungelson}, spectral density of gravitational-wave signal is given by
\be
S_h(2\pi f)=\frac{\pi}{12}\frac{(G{\cal M})^{5/3}}{c^3r^2}\frac{\Theta(f_{\rm max}-f)}{(\pi f)^{7/3}}.
\ee
Here the "chirp" mass $\cal M$ is defined as ${\cal M}\equiv \mu^{3/5}M^{2/5}$ with
$\mu$ and $M$ being the reduced mass and total mass of the binaries system.
With other parameters being fixed, the corresponding spectrum shows a frequency dependence
of $f^{-7/3}$. Therefore, as a measure of the detector sensitivity,
we can defined an integrated signal-to-noise ratio (SNR) for NSNS binaries as
\be
\rho^2_{\rm NSNS}\propto \int_{f_{\rm min}}^{f_{\rm max}}\frac{f^{-7/3}df}{S^{\rm quant}_h(2\pi f)
+S_h^{\rm tech}(2\pi f)}.
\ee
The upper limit of the integral $f_{\rm max}\sim f_{\rm ISCO}\approx 4400\times (M/M_{\bigodot})$ Hz is determined by the innermost stable circular orbit (ISCO) frequency, and the lower limit $f_{\min}$ is set to be $10$ Hz, at which the noise can no longer be considered as stationary. Here we choose $M=2.8 M_{\bigodot}$, the same as in Ref. \cite{Khalili5}.
Here $S_{h}^{\rm quant}$ is the quantum noise spectrum derived in the previous sections and $S_{h}^{\rm tech}$ corresponds to the technical noise obtained
from {\it Bench} \cite{Bench}.

Another interesting astrophysical sources are Bursts \cite{Abbott}.
The exact spectrum is not well modeled and a usual applied simple model
is to assume a logarithmic-flat signal spectrum, i.e. $S_h(2\pi f)\propto f^{-1}$.  The corresponding integrated SNR is then given by
\be
\rho^2_{\rm Bursts}\propto \int_{f_{\rm min}}^{f_{\rm max}}\frac{d\log f}{S^{\rm quant}_h(2\pi f)
+S_h^{\rm tech}(2\pi f)}.
\ee
The integration limit is taken to be the same as the NSNS binaries case.

To estimate the SNR and also motivate future implementation of this scheme, we assume the filter cavity
has a length of $\sim 30$ m and an achievable optical loss 10ppm per bounce and also consider the non-unity
quantum efficiency of the photodiodes $\eta=0.9$ for both {\sf MHD} and {\sf AHD}. Other relevant parameters will be further optimized numerically. For comparison, we will also optimize other related configurations, which includs AdvLIGO, AdvLIGO with frequency-independent squeezed-input (FISAdvLIGO for short), and CMW scheme. Specifically, free parameters for these different schemes that need to be optimized are the following:
\begin{subequations}
  \begin{align}
    &\text{AdvLIGO:} & &r_{\rm SR}\,,\ \phi_{\rm SR}\,,\ \phi\,, \\
    &\text{FISAdvLIGO:} & &r_{\rm SR}\,,\ \phi_{\rm SR}\,,\ \phi\,,\ \theta\,, \\
    &\text{CMW:} & &r_{\rm SR}\,,\ \phi_{\rm SR}\,,\ \phi\,,\ \gamma_{\rm\sf I},\ \gamma_{\rm\sf E},\ \theta_s\,,\theta_p, \\
    &\text{CMWA:} & &r_{\rm SR}\,,\ \phi_{\rm SR}\,,\ \phi\,,\ \gamma_{\rm\sf I},\ \gamma_{\rm\sf E},\ \theta_s\,,\theta_p\,, \zeta\,.
  \end{align}
\end{subequations}

\begin{table}[!ht]
\caption{Optimization results for different schemes}\label{tab:results}
\renewcommand{\arraystretch}{0.8}
  \begin{tabular}{|c|c|c|c|c|c|c|c|c|c|c|c|c|c|c|c|c|}
    \hline
    & \multicolumn{8}{c|}{\bf NSNS} & \multicolumn{8}{c|}{\bf Bursts} \\
    \cline{2-17}
    \raisebox{2ex}[0pt]{\bf Configurations}& $\rho$
    & $r_{\rm SR}$ & $\phi_{\rm SR}$ & $\gamma_{\rm\sf I}$ & $\gamma_{\rm\sf E}$&$\phi$ &$\theta_p$ & $\zeta$
        & $\rho$ & $r_{\rm SR}$ & $\phi_{\rm SR}$ & $\gamma_{\rm\sf I}$ & $\gamma_{\rm\sf E}$ &$\phi$ &$\theta_p$& $\zeta$ \\
    \hline
      AdvLIGO                   & 1.0 & 0.8 & 1.4        & --- & --- & $-1.0$ & --- &--- &1.0 & 0.7 & 1.5 & --- & ---& $-0.2$ & --- & ---\\
      FISAdvLIGO                & 1.0 & 0.7 & 1.5        & --- & --- & $-1.2$ & $-0.6^{\S}$ &---&1.5 & 0.8 & $1.6$ &---&---& 0.0 & $-1.6^{\S}$ &---\\
      CMW                       & 1.0 & 0.8 & $1.6$ & 240 & 240 & $-0.8$ &$-0.6$ &--- & 1.5 & 0.8 & $1.6$  & 0.0   & 0.0 &0.0&$-1.4$&---\\
      CMWA                      & 1.2 & 0.7 & $1.6$ & 230 & 210 & 0.0 &$-0.1$ & 0.7 & 1.5 & 0.8 & $1.6$  & 140 & 140 &0.0&$-0.2$& 0.9\\
    \hline
  \end{tabular}
  \newline\noindent $^{\S}$ This is the squeezing angle $\theta$ in the case of single squeezed-input.
\end{table}

The resulting optimal parameters for different schemes are listed in Table \ref{tab:results}.
They are rounded to have two significant digits at most to balance various uncertainties
in the technical noise. The integrated SNR $\rho$ is normalized with respect to that of the AdvLIGO configuration. The optimal
$\theta_s$ from the numerical result is in accord with the asymptotic estimation, namely $\theta_s\approx(\pi/2)-\phi$ (c.f. Eq. \eqref{thetasopt}).

\begin{figure}
\centerline{\includegraphics[width=0.5\textwidth, bb=6 15 328 227,clip]{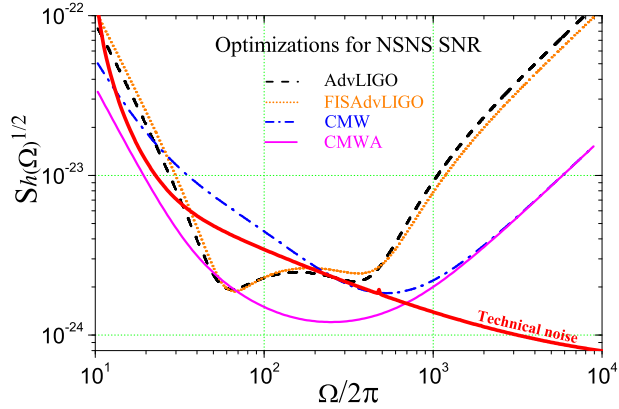}}\caption{
Quantum-noise spectrums of different schemes with optimized parameters
for detecting gravitational waves from NSNS binaries. The optimal values for the parameters are
listed in Table. \ref{tab:results}.}
\label{nsnsopt}
\end{figure}

The corresponding quantum-noise spectrums of different
schemes optimized for detecting gravitational waves from NSNS binaries are shown in Fig. \ref{nsnsopt}.
The FISAdvLIGO, FISAdvLIGO and CMW schemes almost have the identical integrated sensitivity and
CMWA scheme shows moderate $20\%$ improvement in SNR. This is attributable to the fact
that the signal spectrum of NSNS binaries has a $f^{-7/3}$ dependence and low-frequency
sensitivity is very crucial. However, due to low-frequency technical
noise, advantages of the CMWA scheme are at most limited in the case for detecting low-frequency sources.

\begin{figure}
\centerline{\includegraphics[width=0.5\textwidth, bb=6 15 328 227,clip]{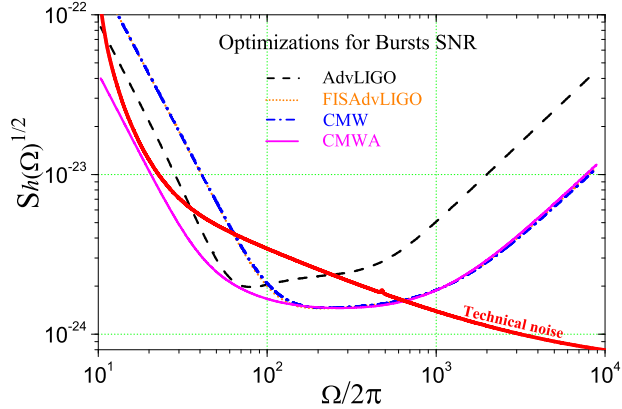}}\caption{
Quantum-noise spectrums of different schemes which are optimized
for detecting gravitational waves from Bursts. The optimal values for the parameters are listed in Table. \ref{tab:results}.}
\label{burstopt}
\end{figure}

The case for detecting gravitational waves from Bursts is shown in Fig. \ref{burstopt}. All other three
schemes have a significant $50\%$ improvement in terms of SNR over AdvLIGO. The sensitivities
of the optimal FISAdvLIGO, CMW and CMWA at high frequencies almost overlap each other.
In addition, detuned phase $\phi_{\rm SR}$ of the signal recycling cavity of those
three are nearly equal to $\pi/2$, which significantly increases the effective
detection bandwidth of the gravitational-wave detectors and is the same as the resonant-sideband
extraction scheme. This is because a broadband sensitivity is
preferable in the case of Bursts which have a logarithmic flat spectrum.

\begin{figure}
\includegraphics[width=0.5\textwidth, bb=6 15 328 227,clip]{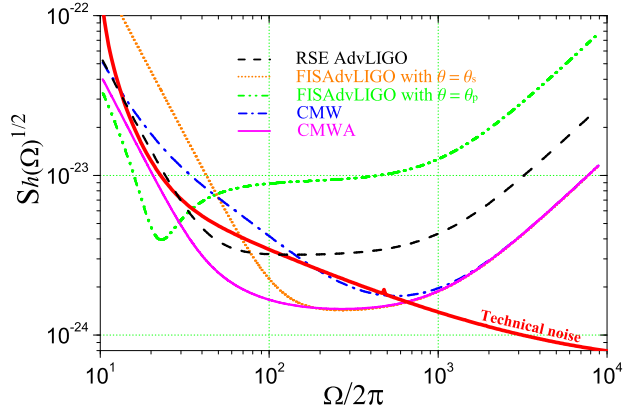}\caption{
Quantum-noise spectrums of different schemes using the same parameters as the optimal CMWA to show how
different parameters affect sensitivity of the CMWA scheme. }
\label{comp}
\end{figure}

To show explicitly how different parameters affect sensitivity of the CMWA scheme,
we present quantum-noise spectrums of different schemes using the same parameters as the optimal CMWA in Fig. \ref{comp}.
In the case of AdvLIGO, we obtain a resonant-sideband extraction (RSE) configuration with
$\phi_{\rm SR}\approx\pi/2$. The quantum noise of FISAdvLIGO with squeezing angle
$\theta=\theta_s$ is lower at high frequencies but higher at low frequencies than the RSE AdvLIGO. FISAdvLIGO with $\theta=\theta_p$ behaves in the opposite way with significant increase of sensitivity
at low frequencies but worse at high frequencies.
The CMW scheme with double squeezed-input, just as expected,
can improve the sensitivity at both high and low frequencies but performs not so well at the intermediate
frequencies. The CMWA scheme performs very nice over the whole observational band compared with others and it would be more attractive if the technical noise of the AdvLIGO design can be further decreased.

\section{Conclusion}\label{IV}

We have proposed and analyzed the double squeezed-input CMWA scheme as an option for
increasing sensitivity of future advanced gravitational-wave detectors. Given an
achievable optical loss of the filter cavity and 10dB squeezing,
this CMWA configuration shows a noticeable reduction in quantum noise at both
high and low frequencies compared with other schemes. Since the length of the filter
cavity considered here is around $30$ m, with the developments of low-loss
coating and better squeezing-state sources, it could be a promising and
relatively simple add-on to AdvLIGO without a need for dramatically modifying
the existing topology.

\begin{center}
{\bf ACKNOWLEDGEMENTS}
\end{center}

We thank S.L. Danilishin for invaluable discussions. F. Ya. Khalili's research has been
supported by NSF and Caltech Grant No. PHY-0651036. H. Miao's research was supported by
the Australian Research Council and the Department of Education, Science and Training.
Y. Chen's research was supported by the Alexander von Humboldt Foundation's Sofja
Kovalevskaja Programme, NSF grants PHY-0653653 and PHY-0601459, as well as the David and Barbara Groce
startup fund at Caltech. H. Miao would like to thank D. G. Blair, J. Li and C. Zhao
for their keen supports of his visiting to the Max Planck Institute and Caltech.

\end{document}